# Secure Polarization-Shift Backscatter Identification Applied to Battery-Free BLE Sensors Powered by Wireless Power Transfer

Taki E. Djidjekh
LAAS-CNRS, Université de Toulouse, CNRS
taki.djidjekh@laas.fr

Quentin Bernyer
LAAS-CNRS, Université de Toulouse, CNRS
quentin.bernyer@laas.fr

Alexandru Takacs
LAAS-CNRS, Université de Toulouse, CNRS, UPS
alexandru.takacs@laas.fr

***Abstract*—This paper presents a lightweight and protocol-independent security mechanism for battery-free Bluetooth Low Energy (BLE) sensor nodes operating in Simultaneous Wireless Information and Power Transfer (SWIPT) architecture. The proposed approach exploits polarization-shift backscattering of the wireless power wave to transmit an encrypted device identification prior to data communication. A fail-safe RF switch and orthogonally polarized antennas are integrated as an external add-on module, enabling controlled backscatter without modifying the original energy-harvesting rectifier. The identification payload is encrypted using AES-128 and transmitted with minimal energy overhead. Experimental validation on a battery-free BLE sensor node demonstrates reliable extraction of the backscattered identification signal, seamless coexistence with BLE advertising, and improved RF-to-DC harvesting efficiency compared to rectifier-based backscatter solutions. The results confirm that polarization-shift backscatter identification provides an effective and practical security for battery-free BLE sensing systems.**

## I. INTRODUCTION

Battery-Free wireless Sensor Nodes (BFSNs) are gaining increasing attention as a sustainable solution for long-term sensing in resource-constrained and hard-to-access environments. By eliminating batteries, these systems reduce maintenance costs and enable continuous operation in scenarios where periodic replacement is impractical. A key enabling technology for such nodes is Wireless Power Transfer (WPT), which provides a predictable and controllable energy source, in contrast to ambient energy harvesting techniques that suffer from intermittency and limited availability. Within this context, Simultaneous Wireless Information and Power Transfer (SWIPT) has emerged as a promising framework for jointly supporting energy delivery and wireless communication [1], [2].

SWIPT-based BFSNs typically rely on either single-wave or dual-wave architectures [3], [4]. In single-wave systems, the same RF signal is used for both energy harvesting and backscatter communication, resulting in highly energy-efficient but functionally constrained nodes. Dual-wave architectures decouple power delivery and data communication, allowing BFSNs to operate onboard microcontrollers, interface with low-cost sensors, and transmit data using standardized protocols such as Bluetooth Low Energy (BLE) or LoRaWAN. This increased functionality, however, comes at the cost of extremely limited energy budgets for security-related operations.

Due to these constraints, BFSNs often employ minimal or no cryptographic protection. In BLE-based battery-free systems, energy limitations frequently lead to connectionless advertising without pairing or authentication, exposing the network to spoofing, replay, and denial-of-service attacks [5]. Existing SWIPT security solutions, including jamming-based methods [6], spread-spectrum techniques [7], intelligent reflecting surfaces [8], and deep learning-based approaches [9], are generally hardware-intensive, protocol-specific, needs accurate channel state information, or computationally demanding, making them unsuitable for ultra-low-power battery-free platforms.

In this paper, a promising physical-layer identification mechanism based on backscattering the WPT power wave is extended to enable secure and protocol-agnostic device authentication. A physical layer identification mechanism was previously demonstrated using a dedicated backscattering rectifier to modulate an identification signal on the power wave [10]. Subsequently, this identification principle was implemented through a polarization-shift rectenna add-on module composed of a RF switch connected with an efficient energy-harvesting rectifier [11], eliminating the need for a backscattering rectifier . Building on this latter approach, the solution proposed in this paper employs an external add-on module composed of a fail-safe RF switch and two antennas with orthogonal polarizations. Controlled switching between these antennas enables low-rate backscatter modulation of an identification signal while maintaining efficient wireless power transfer.

While previous validation was limited to a LoRaWAN-based BFSN, this work experimentally confirms that polarization-shift-based backscatter identification can be seamlessly combined with BLE communication without affecting normal operation. In addition to functional validation, extensive experimental characterizations and comparative measurements are conducted to evaluate identification reliability, robustness, and energy efficiency under BLE operation. The results highlight the protocol-independent, energy-efficient, and non-intrusive nature of the proposed mechanism, supporting its suitability for practical deployment in battery-free BLE sensing systems.

The remainder of the paper is organized as follows. Section II describes the WPT-based identification principle. Section III presents the experimental validation using a battery-free BLE sensor prototype incorporating the add-on polarization-shift rectenna module. Section IV discusses the experimental results, and Section V concludes the paper and outlines future perspectives.

## II. WPT-Based Polarization-Shift Identification Principle

In a dual-wave SWIPT system, wireless power delivery and data transmission are handled by two distinct RF signals. The power wave (P-wave), emitted by the Communication Node (CN), continuously supplies energy to the BFSN, while a separate communication wave supports data exchange via the BFSN transceiver. The identification mechanism considered in this work exploits the P-wave beyond its conventional role of energy provisioning by enabling device identification before the data transmission.

The BFSN modulates an identification pattern onto the incident P-wave through controlled backscattering while the data transceiver is inactive, allowing authentication to take place with negligible energy overhead. This modulation is driven by a digital private key (PvK) generated by the onboard ultra-low-power microcontroller, which controls the backscatter operation using simple GPIO-based switching, as illustrated in Fig. 1. As a result, identification can be performed without activating the RF transceiver, significantly reducing energy consumption.

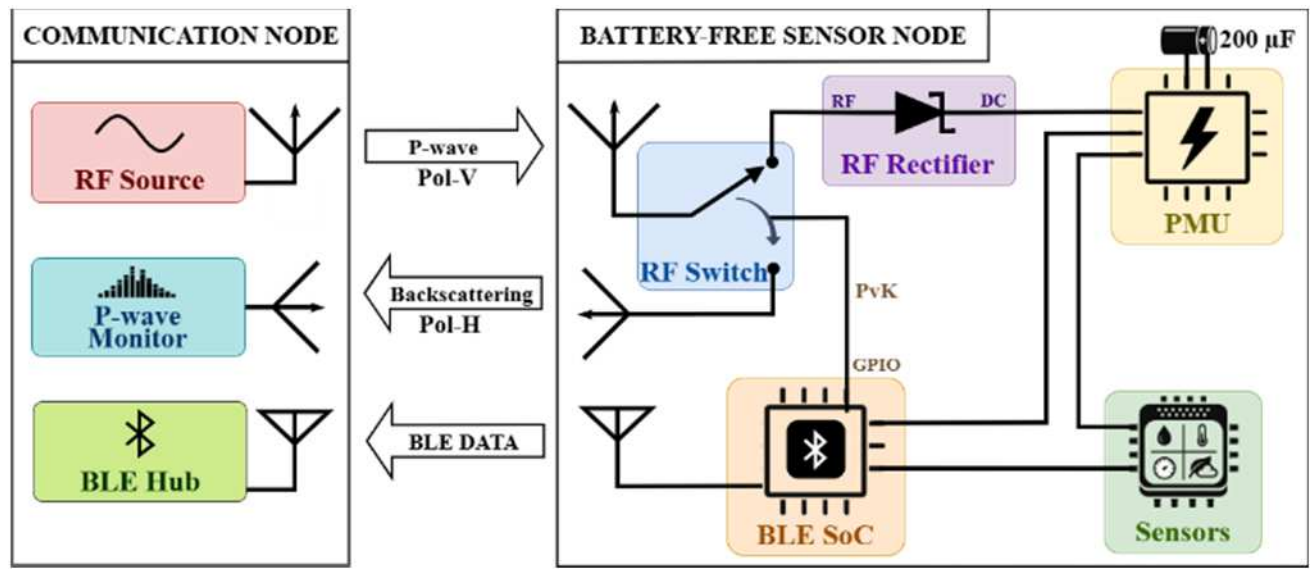

Fig. 1. Dual-wave SWIPT architecture with a polarization-shift backscatter security layer, highlighting the add-on module at the BLE battery-free node and the P-wave monitor at the communication node.

At the CN, where power and computational resources are not constrained, a dedicated P-wave monitoring receiver observes the uplink backscattered signal. Since a single CN can simultaneously energize and manage multiple BFSNs, this monitoring functionality enables scalable and centralized identification verification. A defining aspect of the proposed approach is that the backscattered identification signal is transmitted using a polarization orthogonal to that of the incident P-wave, as shown in Fig. 1 This polarization shift reduces the impact of environmental reflections, multipath clutter, and cross-jamming, thereby improving the robustness of identification in practical deployment scenarios.

In this work, the concept is implemented on a BLE-based battery-free sensor node built around the NXP QN9080 BLE SoC and a 2.45 GHz PIFA antenna. The node integrates a power-management unit (AEM30940, e-peas) with a 220 µF storage capacitor and an 868 MHz RF rectifier, and uses a temperature-humidity sensor to report sensing data.

The add-on module consists of a fail-safe SPDT RF switch (GRF6011, Guerrilla RF) connected to two 2.5 dBi monopole antennas with orthogonal linear polarizations (V and H). With no supply or control signal, the switch stays in its default state, routing the BFSN rectifier to the V-polarized antenna with a low insertion loss (≤ 0.4 dB) at the RF input. Once sufficient energy is available, the QN9080 toggles the switch to route the incident P-wave to the H-polarized antenna, enabling controlled backscatter in the orthogonal polarization according to a digital private key (PvK). The PvK frame includes a 16-bit preamble (0xAAAA) followed by a 16-byte key, Manchester-coded at up to 50 kHz. The key is generated using AES-128, with variability introduced from internal ADC measurements.

The BLE-based BFSN measures temperature and humidity after each energy recharge cycle and subsequently advertises the sensed data without commissioning. For improved robustness, the advertising procedure is repeated four times over different BLE advertising channels. Prior to each BLE advertising event, the PvK sequence is backscattered on the WPT P-wave to provide a complementary device identification. This identification step enhances security against replay and flooding attacks by enabling authentication of the BFSN before data transmission.

## III. Experimental Validation on a BLE BFSN

### A. Backscattering Identification Test

An experimental platform was set up to assess the polarization-shift backscatter approach under realistic conditions. The setup comprised the BLE-based BFSN equipped with the add-on module and a CN. The CN combined an RF power source and a Tektronix RSA306B USB spectrum analyzer for observation of the backscattered component, alongside a BLE hub used to collect the sensor data through conventional BLE reception, as depicted in Fig. 2.

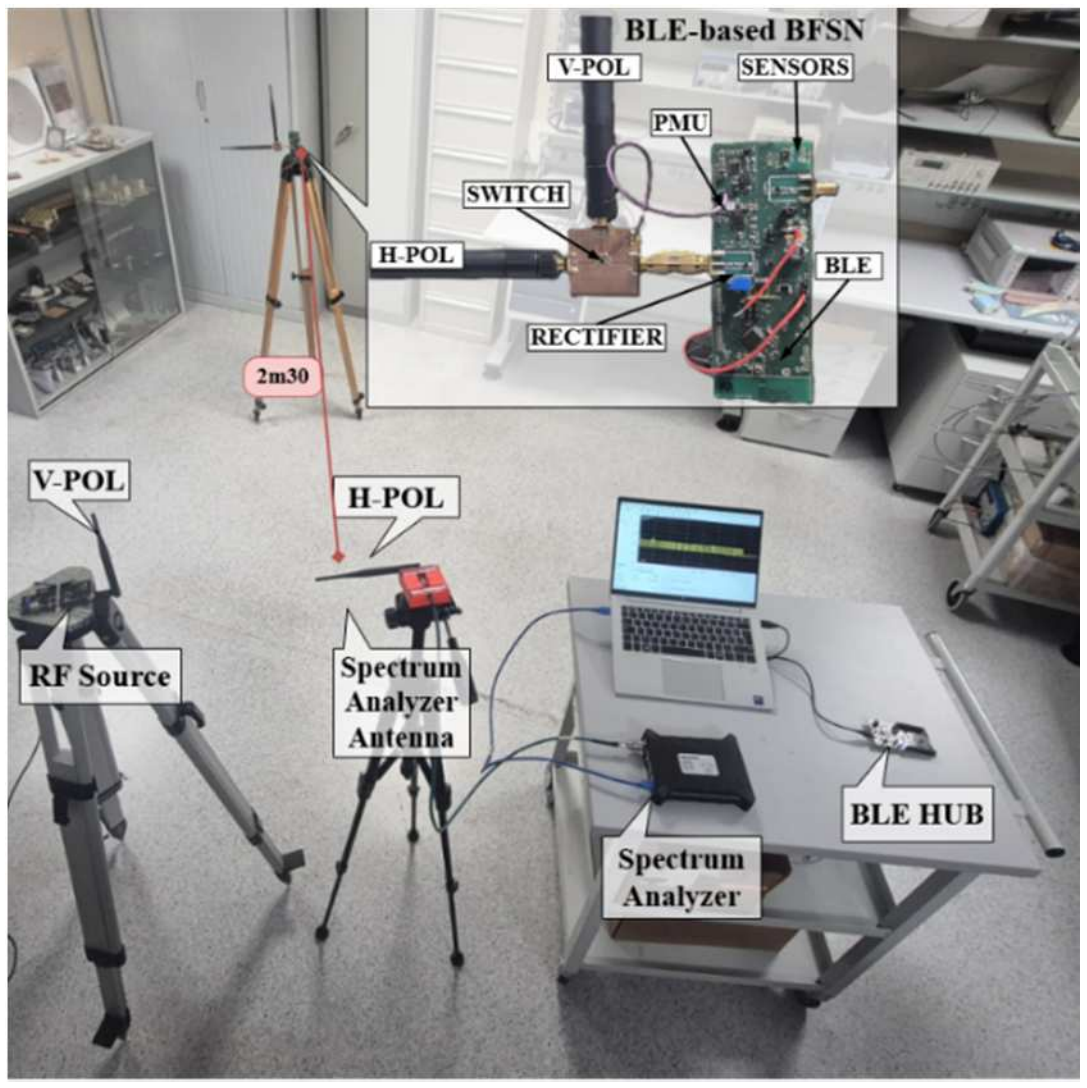

Fig. 2. Experimental setup illustrating the Communication Node (including the RF power source, BLE hub, and spectrum analyzer) and the battery-free sensor node equipped with the fail-safe switch, with orthogonal polarizations assigned to the incident P-wave and the backscattered signal.

At the CN, the RF source fed a vertically polarized monopole antenna identical to the BFSN antennas (gain: 2.5 dBi), whereas the spectrum analyzer was connected to an identical monopole configured for horizontal polarization. The two CN antennas were therefore orthogonally polarized and arranged to reduce leakage from the transmit chain into the receive chain by aligning the radiation minimum of the horizontally polarized antenna toward the vertically polarized transmitting antenna. In this arrangement, vertical polarized waveforms support efficient WPT delivery to the BFSN harvesting antenna, while the horizontally polarized waveforms are used to backscatter secure authentication from the BFSN to CN.

For the measurements, the BFSN was placed 2.3 m from the CN (Fig. 2). The RF source continuously emitted an 868 MHz P-wave with an EIRP of 33 dBm, remaining compliant with European transmission regulations.

The BFSN backscattered the PvK by toggling the fail-safe RF switch between the V-polarized harvesting antenna and the H-polarized backscatter antenna, thereby modulating the reradiated signal via polarization switching. The Manchester-coded signature was detected by the spectrum analyzer before each BLE advertising event. As shown in Fig. 3, the PvK was reliably recovered and verified at the CN, providing an added authentication layer without modifying the original BFSN circuitry.

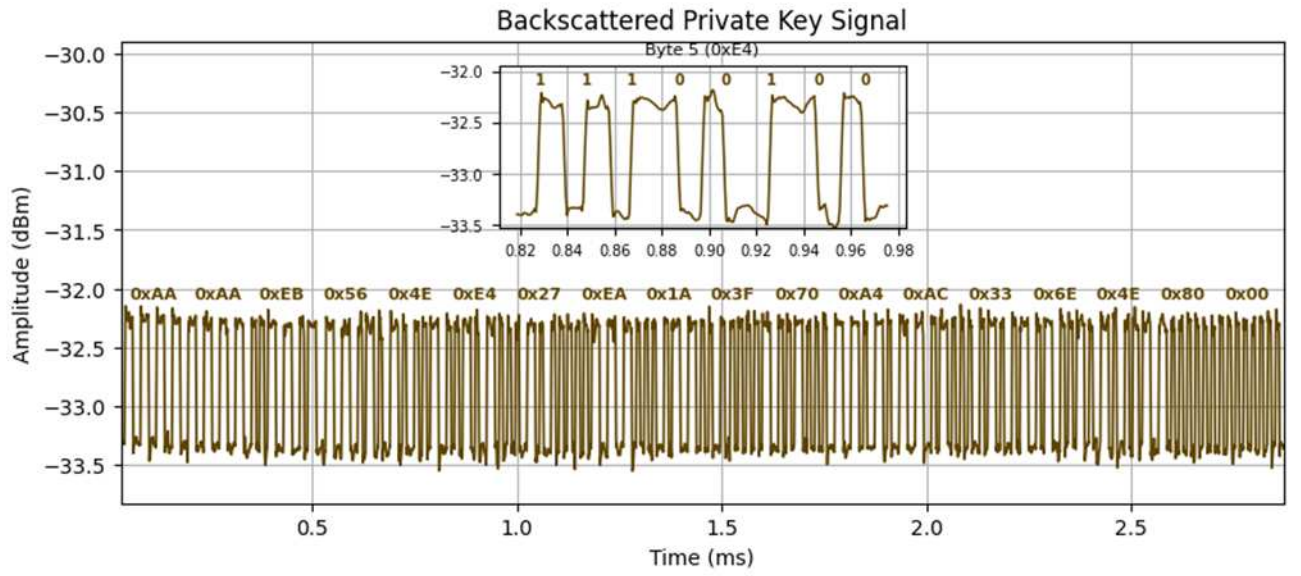

Fig. 3. Polarization-shift backscattered private key captured by the spectrum analyzer, showing the complete 18-byte sequence (16-bit preamble 0xAAAA followed by a 16-byte AES-128–encrypted PvK), with an inset highlighting the bit-level structure of a single byte.

## B. *RF-to-DC Efficiency and Power Consumption*

To enable a fair and quantitative comparison between the proposed add-on module (fail-safe SPDT switch with orthogonally polarized antennas) and backscattering-rectifier-based approaches [12], the RF-to-DC conversion efficiency was experimentally characterized for three distinct configurations: (i) The original BFSN harvesting rectifier, illustrated in Fig. 4, used as a reference for optimal harvesting performance. (ii) A dedicated backscattering rectifier inspired by [10], designed to simultaneously support energy harvesting and backscattering. (iii) The proposed add-on module configuration, using the original rectifier with the fail-safe switch.

Measurements were carried out at 868 MHz, the common best-match operating point for the three prototypes, using a 10 kΩ resistive load while sweeping the applied RF input power. This methodology enables a direct evaluation of how each backscatter solution affects energy-harvesting performance; the corresponding results are presented in Fig. 5.

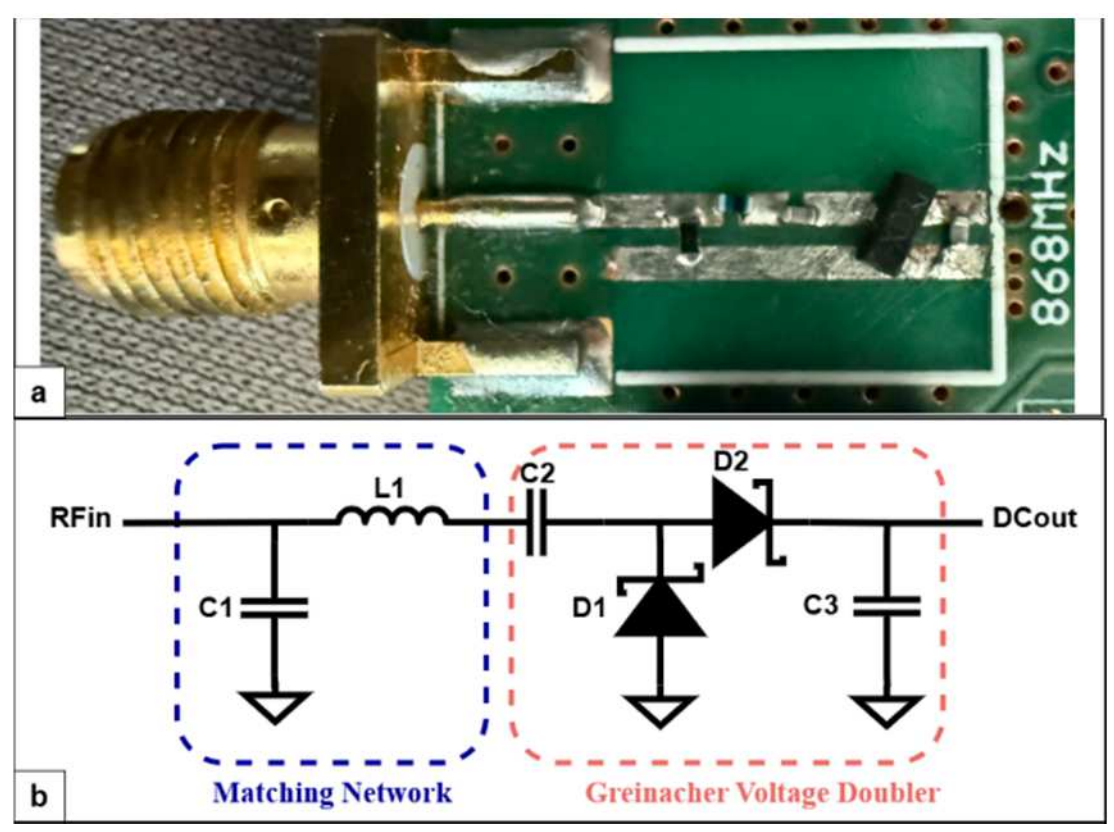

Fig. 4. (a) Photograph of the 868 MHz BFSN RF rectifier; (b) corresponding schematic with SMS7630-005LF Schottky diode pair (D1, D2), C1 = 4 pF, and L1 = 33 nH (LQW15AN33NG00).

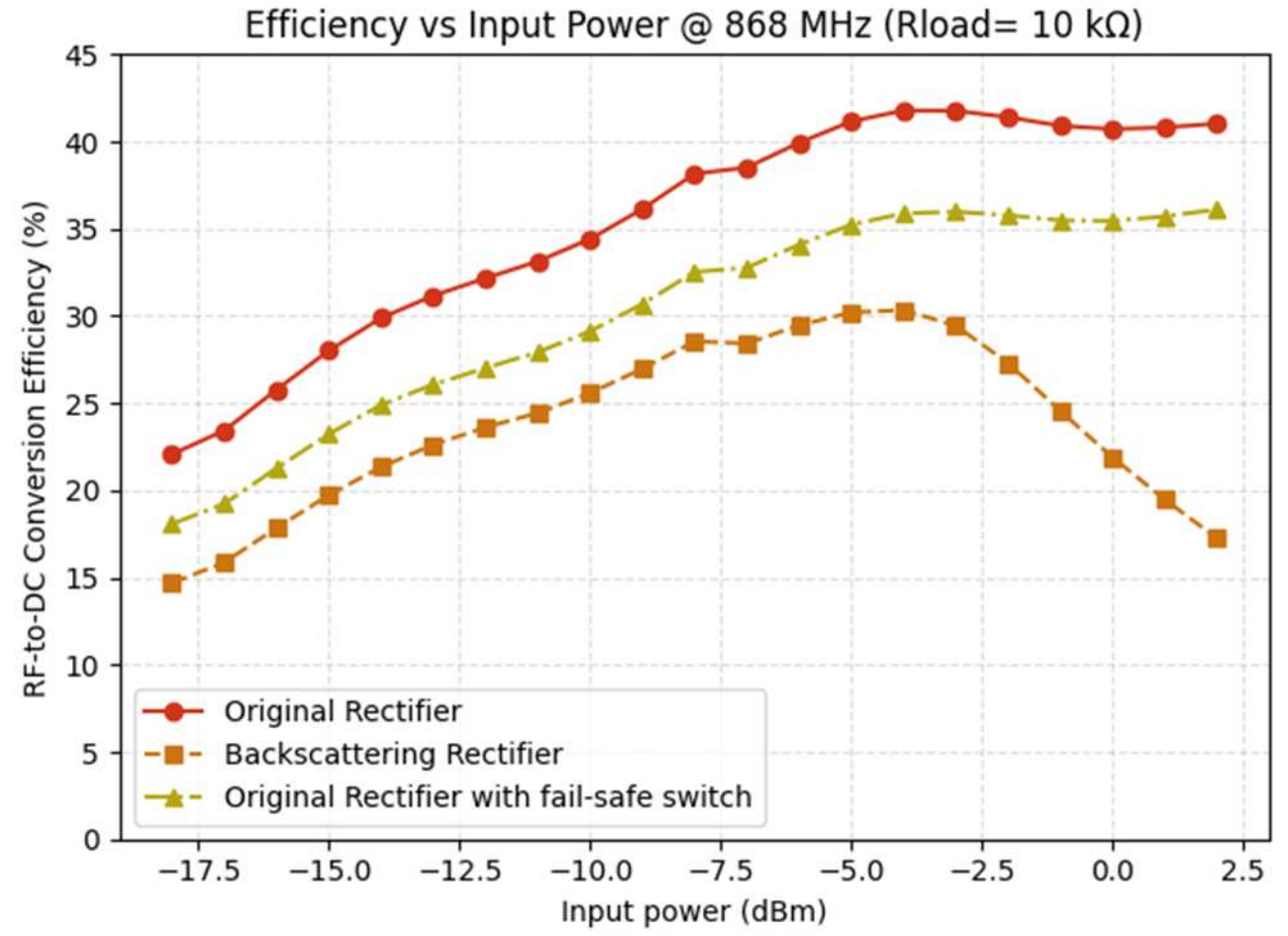

Fig. 5. RF-to-DC conversion efficiency versus input RF power at 868 MHz with a 10 kΩ load, comparing the original BFSN rectifier, a backscattering rectifier, and the proposed fail-safe switch interface driving the original rectifier.

Fig. 5. shows that both backscatter solutions reduce RF-to-DC efficiency relative to the original rectifier, due to added losses. Despite this, the proposed add-on module (≤ 0.4 dB insertion loss) exceeds the backscattering rectifier by at least 3% over the full input-power sweep and tracks the baseline rectifier trend, indicating preserved harvesting behavior. At higher input powers, it delivers >15% higher efficiency than the backscattering rectifier, while requiring no modification to the BFSN circuitry.

After confirming identification backscatter operation and harvesting efficiency, the energy overhead of the identification step was quantified. In original mode, the BFSN runs a periodic cycle including initialization, sensor reading, and BLE advertising events. The secured mode inserts a 3 ms GPIO-driven backscatter sequence immediately before each advertising events. Current profiles were measured for the three cases: (i) original BFSN, (ii) identification via a backscattering rectifier, and (iii) identification via the proposed fail-safe RF-switch add-on. The resulting time-domain current consumption traces enable direct comparison of the overhead introduced by each option.

As illustrated in Fig. 6, both identification schemes introduce a temporal shift in the current consumption profile compared to the original BFSN operation, due to the insertion of the ~3 ms backscattering sequence. The backscattering-rectifier approach increases energy consumption by approximately 52 µJ, while the fail-safe RF-switch add-on module results in an overhead of 93 µJ, corresponding to an additional 41 µJ compared to the backscattering rectifier-based method. Despite this difference, the added energy remains small relative to the total energy per operating cycle. These results indicate that the proposed authentication mechanism can be implemented with minimal energy impact, while the switch-based add-on module solution preserves the original BFSN circuitry without hardware modification.

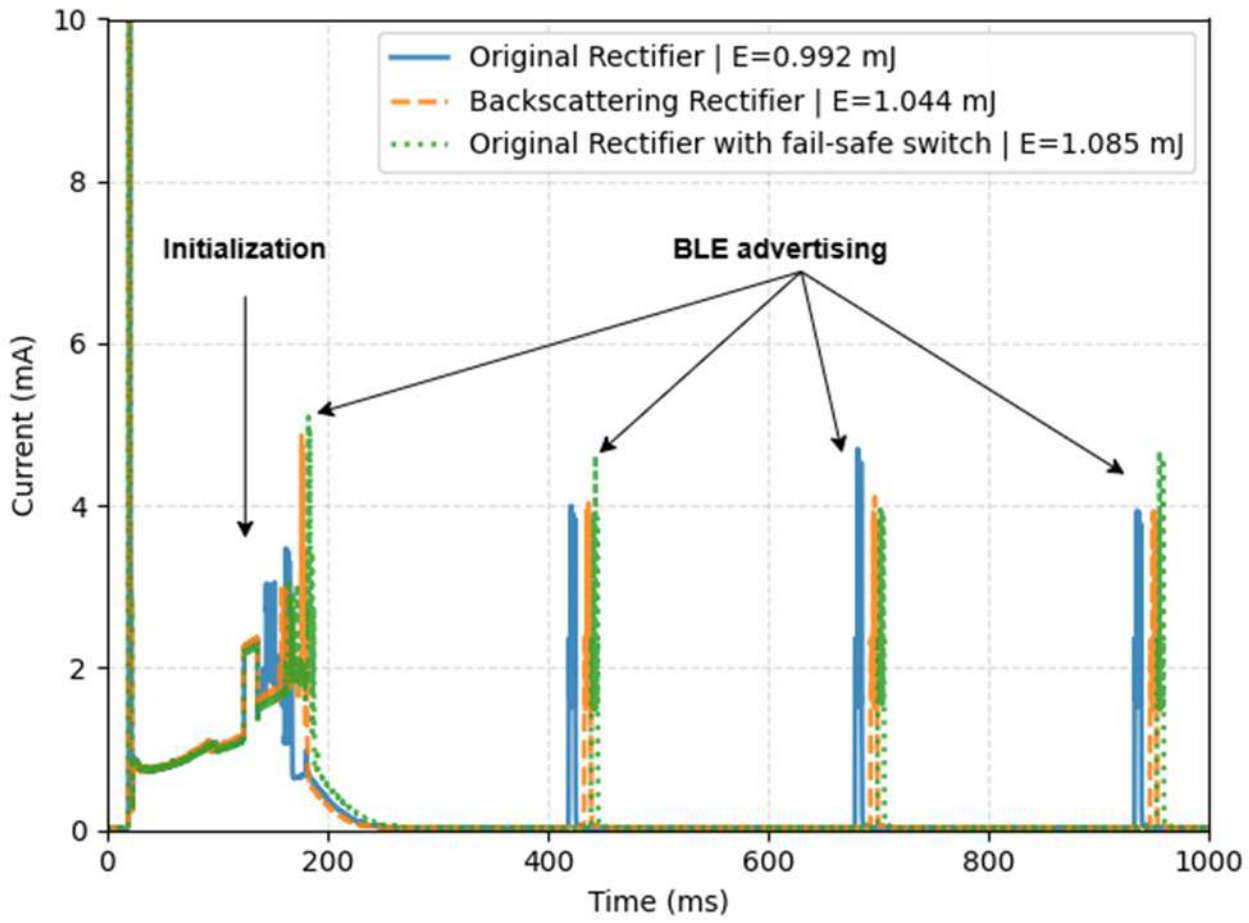

Fig. 6. Measured current consumption versus time over one BFSN operating cycle (initialization, sensing, and 4 BLE advertising), comparing the original node with two identification options: backscattering-rectifier-based modulation and the proposed fail-safe RF-switch add-on inserting a ~3 ms backscatter sequence before each advertising event.

## IV. Discussion

The measurements validate that polarization-shift backscattering controlled by a fail-safe SPDT switch can provide a practical identification layer for a battery-free BLE sensor node. The PvK was consistently recovered at the Communication Node before each BLE advertising event, showing that reliable identification can be performed without a dedicated backscattering rectifier and without altering the original harvesting front-end. Using an orthogonal polarization for the backscattered component improves separability from the incident P-wave and increases robustness against clutter and cross-interference in real environments.

The PvK in this implementation is AES-128 encrypted (16-byte payload), aligning the identification format with widely used BLE/IoT security primitives while maintaining low computational overhead. By combining encrypted PvK backscattering with repeated BLE advertising, the system enables device authentication prior to data transmission at minimal energy cost. This mechanism effectively acts as a lightweight, pre-advertising verification step, functionally similar to a pairing operation, without requiring the conventional BLE pairing procedure, which is typically avoided in battery-free nodes due to energy constraints.

From an energy perspective, the add-on module remains passive during harvesting, introducing only a small insertion loss (≤0.4 dB), and consumes power solely during brief switching intervals. The insertion of the ~3 ms identification sequence before each BLE advertising event results in a limited energy overhead relative to the full recharge-and-transmit cycle consumption, confirming that the added security function is achieved with a low and well-controlled energy cost.

## V. Conclusion

This work demonstrates the feasibility of polarization-shift backscatter identification in a battery-free BLE sensor node using a non-intrusive fail-safe RF-switch add-on. A locally generated private key is backscattered over the WPT link through orthogonal polarization, enabling an independent identification channel without activating the BLE transceiver. The AES-128–encrypted PvK was reliably recovered at the Communication Node prior to each BLE advertising event, establishing a protocol-independent authentication layer.

The proposed approach introduces a limited energy overhead of 93 µJ within a total cycle energy of 1.085 mJ, while preserving the original harvesting architecture and achieving RF-to-DC efficiencies up to 35%, outperforming rectifier-based backscatter solutions. These results confirm that polarization-shift backscatter provides a practical, energy-efficient, and easily integrable security enhancement for battery-free BLE systems.

## Acknowledgement

This research was funded, in whole or in part, by the French National Research Agency (ANR) under the project SWAVE “ANR-25-CE39-5853”, and the authors gratefully acknowledge its support.

## References

[1] R. L. Rosa, P. Livreri, C. Trigona, L. D. Donato, and G. Sorbello, “Strategies and Techniques for Powering Wireless Sensor Nodes through Energy Harvesting and Wireless Power Transfer,” *Sensors*, vol. 19, no. 12, June 2019, doi: 10.3390/s19122660.

[2] G. Loubet *et al.*, “Wirelessly Powered Battery-Free Sensing Nodes for Internet of Things Applications,” *IEEE Microwave Magazine*, vol. 26, no. 7, pp. 26–46, July 2025, doi: 10.1109/MMM.2024.3488593.

[3] Y. Yao, P. Sun, X. Liu, Y. Wang, and D. Xu, “Simultaneous Wireless Power and Data Transfer: A Comprehensive Review,” *IEEE Transactions on Power Electronics*, vol. 37, no. 3, pp. 3650–3667, Mar. 2022, doi: 10.1109/TPEL.2021.3117854.

[4] J. Huang, C.-C. Xing, and C. Wang, “Simultaneous Wireless Information and Power Transfer: Technologies, Applications, and Research Challenges,” *IEEE Communications Magazine*, vol. 55, no. 11, pp. 26–32, Nov. 2017, doi: 10.1109/MCOM.2017.1600806.

[5] D. H. Morais, “Bluetooth 5/6 Overview,” in *5G/5G-Advanced, Wi-Fi 6/7, and Bluetooth 5/6: A Primer on Smartphone Wireless Technologies*, D. H. Morais, Ed., Cham: Springer Nature Switzerland, 2025, pp. 181–204. doi: 10.1007/978-3-031-82830-0_10.

[6] R. Ma, H. Wu, J. Ou, S. Yang, and Y. Gao, “Power Splitting-Based SWIPT Systems With Full-Duplex Jamming,” *IEEE Transactions on Vehicular Technology*, vol. 69, no. 9, pp. 9822–9836, Sept. 2020, doi: 10.1109/TVT.2020.3002976.

[7] P. Gong, T. M. Chen, P. Xu, and Q. Chen, "DS-SWIPT: Secure Communication with Wireless Power Transfer for Internet of Things," *Security and Communication Networks*, vol. 2022, pp. 1–11, June 2022, doi: 10.1155/2022/2650474.

[8] H. T. Thien, P.-V. Tuan, and I. Koo, "A Secure-Transmission Maximization Scheme for SWIPT Systems Assisted by an Intelligent Reflecting Surface and Deep Learning," *IEEE Access*, vol. 10, pp. 31851–31867, 2022, doi: 10.1109/ACCESS.2022.3159679.

[9] V. Ganapathy, R. Ramachandran, and T. Ohtsuki, "Deep Learning Methods for Secure IoT SWIPT Networks," *IEEE Internet of Things Journal*, vol. 11, no. 11, pp. 19657–19677, June 2024, doi: 10.1109/JIOT.2024.3368692.

[10] T. E. Djidjekh *et al.*, "Backscattering Rectifier for Security and Identification in the context of Simultaneous Wireless Information and Power Transfer," in *2024 54th European Microwave Conference (EuMC)*, Paris, France: IEEE, Sept. 2024, pp. 300–303. doi: 10.23919/EuMC61614.2024.10732847.

[11] T. E. Djidjekh and A. Takacs, "Polarization-Shift Backscatter Identification for SWIPT-Based Battery-Free Sensor Nodes," *Electronics*, vol. 15, no. 1, Dec. 2025, doi: 10.3390/electronics15010186.

[12] T. E. Djidjekh, G. Loubet, and A. Takacs, "Backscattering-Based Security in Wireless Power Transfer Applied to Battery-Free BLE Sensors," in *2025 IEEE Wireless Power Technology Conference and Expo (WPTCE)*, Rome, Italy: IEEE, June 2025, pp. 1–4. doi: 10.1109/WPTCE62521.2025.11062233.